\documentclass[a4paper,10pt]{article}

\usepackage{amsmath,amssymb,mathtools}     
\usepackage{color}
\usepackage{graphicx}
\usepackage{subfigure}
\usepackage{cite}                
\usepackage{hyperref}            
\usepackage{multirow,makecell}   
\usepackage{textcomp}
\usepackage{wasysym}

\usepackage[text={17.5cm,25cm},centering]{geometry}

\numberwithin{equation}{section}   

\def \be {\begin{equation}}
\def \ee {\end{equation}}
\def \ba {\begin{array}}
\def \ea {\end{array}}
\def \bea{\begin{eqnarray}}
\def \eea{\end{eqnarray}}
\def \nn {\nonumber}

\def \a {\alpha}
\def \b {\beta}

\def \G {\Gamma}

\def \e {\epsilon}

\def \m {\mu}

\def \s {\sigma}

\def \r {\rho}

\def \t {\tau}
\def \z {\zeta}

\def \mA {\mathcal A}
\def \mB {\mathcal B}
\def \mC {\mathcal C}
\def \mD {\mathcal D}
\def \mE {\mathcal E}

\def \mN {\mathcal N}

\def \mT {\mathcal T}

\def \mW {\mathcal W}

\def \p {\partial}
\def \f {\frac}

\def \mc {\mathcal}

\def \lt {\left}
\def \rt {\right}

\def \td {\tilde}
\def \hs {\hspace}

\def \inf {\infty}

\def \lag {\langle}
\def \rag {\rangle}

\def \ep {\mathrm{e}}
\def \ii {\mathrm{i}}

\def \tr {\textrm{tr}}

\def \and {{\textrm{and}}}

\def \CFT {{\textrm{CFT}}}

\def \vac {{\textrm{vac}}}

\def \cl {{\textrm{cl}}}

\def \oloop {{\textrm{1-loop}}}

\def \Z {{\textrm{Z}}}
\def \SL {{\textrm{SL}}}

\def \cL {{\textrm{L}}}
\def \NL {{\textrm{NL}}}
\def \NNL {{\textrm{NNL}}}

\begin{document}

\title{\textbf{Short interval expansion of R\'enyi entropy on torus}}
\author{Bin Chen$^{1,2,3}$\footnote{bchen01@pku.edu.cn}~,
Jun-Bao Wu$^{4,5,3}$\footnote{wujb@ihep.ac.cn}~
and
Jia-ju Zhang$^{4,5}$\footnote{jjzhang@ihep.ac.cn}
}
\date{}

\maketitle

\vspace{-10mm}

\begin{center}
{\it
$^{1}$Department of Physics and State Key Laboratory of Nuclear Physics and Technology, Peking University,\\5 Yiheyuan Rd, Beijing 100871, P.R.\,China\\\vspace{1mm}
$^{2}$Collaborative Innovation Center of Quantum Matter, 5 Yiheyuan Rd, Beijing 100871, P.R.\,China\\\vspace{1mm}
$^{3}$Center for High Energy Physics, Peking University, 5 Yiheyuan Rd, Beijing 100871, P.R.\,China\\\vspace{1mm}
$^{4}$Theoretical Physics Division, Institute of High Energy Physics, Chinese Academy of Sciences,\\19B Yuquan Rd, Beijing 100049, P.R.\,China\\\vspace{1mm}
$^{5}$Theoretical Physics Center for Science Facilities, Chinese Academy of Sciences,\\19B Yuquan Rd, Beijing 100049, P.R.\,China
}
\vspace{10mm}
\end{center}

\begin{abstract}

  We investigate the short interval expansion of the R\'enyi entropy for  two-dimensional conformal field theory (CFT) on a torus.  We require the length of the interval $\ell$ to be small with respect to the spatial and temporal sizes of the torus. The operator product expansion of the twist operators allows us to compute the short interval expansion of the R\'enyi entropy at any temperature. In particular, we pay special attention to the large $c$ CFTs dual to the AdS$_3$ gravity and its cousins. At both low and high temperature limits, we read the R\'enyi entropies to order $\ell^6$, and find good agreements with holographic results. Moreover, the expansion allows us to read $1/c$ contribution, which is hard to get by expanding the thermal density matrix. We generalize the study to the case with the chemical potential as well.

\end{abstract}

\baselineskip 18pt
\thispagestyle{empty}
\newpage

\tableofcontents


\section{Introduction}

The entanglement entropy characterizes the entanglement between a system and its environment. It could be defined by the von Neumann entropy of the reduced density matrix
\be
S_A=-\tr_A \r_A\log \r_A.
\ee
In practice, it is often more convenient to compute the entanglement entropy by taking the $n \to 1$ limit of the  $n$-th order R\'enyi entropy
\be
S_A=\lim_{n \to 1} S_A^{(n)},
\ee
where the R\'enyi entropy is defined by
\be
S_A^{(n)}=-\f{1}{n-1} \log \tr_A \r_A^n.
\ee
The study of the R\'enyi entropy is interesting on its own, as it encodes the spectral information of the reduced density matrix.

The entanglement entropy in quantum field theory is difficult to compute, due to the fact that there are infinite degrees of freedom in a field theory\cite{Calabrese:2004eu}, as nicely reviewed in \cite{Casini:2009sr}. If the quantum field theory is conformal invariant, there are more tools to use, in particular for a two-dimensional (2D) conformal field theory (CFT) \cite{Calabrese:2009qy}. The standard way of computing the entanglement entropy in a quantum field theory is the replica trick \cite{Callan:1994py}. For a 2D CFT, one could insert the twist operators at the branch points to impose the nontrivial boundary conditions on the fields in applying the replica trick \cite{Calabrese:2004eu,Calabrese:2009qy}. Then the R\'enyi entropy is determined either by the partition function on the resulting Riemann surface, which is generically of higher genus, or equivalently by the correlation function of the twist operators on the original spacetime manifold but in an orbifold CFT.

The R\'enyi entropy on complex plane has been well studied. It is well known that when there is a single interval of length $\ell$,  the R\'enyi entropy has a universal form, which is independent of details of the CFT and depends only on the central charge $c$ \cite{Calabrese:2004eu}
\be
S_n=\f{c(n+1)}{6n} \log \f{\ell}{\e},
\ee
with $\e$ being the UV cutoff. But for the case with two and more intervals, the R\'enyi entropy would depend on the details of the CFT \cite{Calabrese:2009ez,Headrick:2010zt,Calabrese:2010he}. For the case of two disjoint intervals there have been some studies. The analytical results of R\'enyi entropy $S_n, n\geq2$ for a free compactified  boson and Ising model  have been presented in \cite{Calabrese:2009ez} and \cite{Calabrese:2010he} respectively.
For a general CFT, it was proposed in \cite{Headrick:2010zt} that one can use the operator product expansion (OPE) of the twist operators to compute the R\'enyi entropy $S_n, n\geq2$. This proposal was generalized to find the leading term of R\'enyi entropy with small cross ratio in \cite{Calabrese:2010he}.
This method was also generalized to higher dimensions in \cite{Cardy:2013nua}. More interestingly, the proposal has been applied to the large $c$ CFT with a gravity dual. In \cite{Chen:2013kpa,Chen:2013dxa,Perlmutter:2013paa,Chen:2014kja,Zhang:2015hoa}, the R\'enyi mutual information has been computed and reorganized by the powers of $1/c$.\footnote{If one takes the $n\to1$ limit and is only interested in the von Neumann mutual information, the calculation would be much simpler, as shown in \cite{Beccaria:2014lqa,Li:2016pwu}.}
It was found that the leading order terms agree exactly with the holographic R\'enyi entropy (HRE) \cite{Ryu:2006bv,Ryu:2006ef,Krasnov:2000zq,Faulkner:2013yia}, the next-to-leading terms are in exact agreement with the one-loop correction to  HRE\cite{Barrella:2013wja,Giombi:2008vd}, and moreover there are nonvanishing next-to-next-to-leading terms which should be identified with the two-loop corrections to HRE\cite{Yin:2007gv,Headrick:2015gba}. The study suggests that the holographic computation should be correct even beyond classical gravity, as shown in recent study\cite{Chen:2015uga} of the one-loop partition function of any handle-body configuration from dual CFT.

Another interesting case is the R\'enyi entropy on a torus. Now the CFT is in a thermal state and within a finite size. There are both thermal and finite-size effects in the entanglement entropy. In \cite{Cardy:2014jwa}, the universal thermal correction to the single interval entanglement entropy has been found.  In \cite{Datta:2013hba,Chen:2015cna,Chen:2014hta,Liu:2015iia}, the R\'enyi entropy of free boson on a torus has been discussed.
In \cite{Azeyanagi:2007bj,Headrick:2012fk,Herzog:2013py,Lokhande:2015zma}, the R\'enyi entropy of free fermion on a torus has been studied. In \cite{Schnitzer:2015ira}, the R\'enyi entropy for the $\widehat{\mathrm{SU}}(N)_1$ Wess-Zumino-Witten model on the torus has been discussed. In \cite{Guo:2015uwa}, the thermal effect to the R\'enyi entropy of locally excited states has been discussed.
In particular, the single interval R\'enyi entropy on a torus in the large $c$ CFT dual to the AdS$_3$ gravity has been investigated in \cite{Chen:2014unl,Chen:2015uia,Chen:2015kua} and the exact agreement up to the next-to-leading order with the holographic computation has been found.

The studies in \cite{Cardy:2014jwa,Chen:2014unl,Chen:2015uia,Chen:2015kua} rely on the expansion of the thermal density matrix at the low temperature or at the high temperature.
In the low temperature case, the thermal density matrix could be expanded
\be
\rho=\frac{e^{-\b H}}{\tr e^{-\beta H}}=\frac{1}{\tr e^{-\beta H}} \sum_\phi |\phi\rangle\langle\phi| e^{-\beta E_{\phi}},
\ee
where the summation is over all the normalized and orthogonalized excitations in the theory. For the large $c$ CFT, the contribution from the vacuum Verma module may dominate\cite{Hartman:2013mia} such that one can focus just on the states in the vacuum module. Then the computation is reduced to the multi-point functions on a cylinder. This treatment could be applied to the interval of almost any length, even though in the large interval limit one needs the expansion with respect to the twist sector\cite{Chen:2014hta,Chen:2015kua}. The disadvantage in the treatment is that one has to work in the low temperature or high temperature limit, and the computation becomes complicated beyond the first few excited states, and it is hard to read the $1/c$ terms.

In this work, we would like to apply another strategy to compute the single interval R\'enyi entropy on torus. We focus on the case that the interval is short so that we can apply the OPE of the twist operators. In this way, the computation boils down to the one-point functions of various operators on a torus. Such a strategy could be applied to any CFT one is interested in. Here we pay more attention to the large $c$ CFT. One advantage of this approach is that it works for any temperature, or more generally for any modulus of the torus. By expanding further in the low temperature and high temperature limits, we find consistent picture with the results in \cite{Chen:2014unl,Chen:2015uia}. Furthermore, this new approach allows us to read the $1/c$ terms easily.

The remaining parts of the paper are organized as follows. In Section~\ref{SecRenyi}, we introduce the general framework of our treatment. In Section~\ref{SecVac}, we focus on the vacuum conformal family of a large $c$ CFT and discuss the low and high temperature limits, as well as the effect of chemical potential. In Section~\ref{SecW23} we consider a large $c$ CFT with $\mW(2,3)$ symmetry. In Section~\ref{SecSUSY} we consider a large $c$ $\mN=(1,1)$ superconformal field theory (SCFT). We conclude with some discussions in Section~\ref{SecConc}. In Appendix~\ref{AppBasics} there are some useful CFT basics. In Appendix~\ref{AppLow} there is an alternative calculation of low temperature expansion of the one-point functions (\ref{onepointfunctions}).

\section{R\'enyi entropy on torus}\label{SecRenyi}

In this section, we introduce the general strategy to compute the single interval R\'enyi entropy on a torus. We consider the case that the interval is so short that we may first use the OPE of the twist operators and then sum over the expectation values of one-point functions on the torus. We first briefly review the OPE of the twist operators, and then calculate the single interval R\'enyi entropy on torus.

\subsection{OPE of twist operators}

To calculate the $n$-th order R\'enyi entropy using the replica trick one has to calculate the partition function of the CFT on a Riemann surface, or equivalently one can calculate the correlation functions of twist operators $\s$, $\td\s$ in an orbifold CFT \cite{Calabrese:2004eu}, which is denoted by $\CFT^n$. The twist operators have conformal weights
\be
h_\s=h_{\td\s}=\bar h_\s=\bar h_{\td\s}=\f{c(n^2-1)}{24n}.
\ee
In $\CFT^n$ one has OPE of the twist operators \cite{Headrick:2010zt,Calabrese:2010he,Chen:2013kpa}
\be \label{ope}
\s(z,\bar z)\td \s(0,0)
=\f{c_n}{z^{2h_\s}\bar z^{2\bar h_\s}} \sum_K d_K \sum_{r,s\geq0} \f{a_K^r}{r!}\f{\bar a_K^s}{s!}
                                                                  z^{h_K+r}\bar z^{\bar h_K+s}
                                                                  \p^r \bar \p^s \Phi_K(0,0).
\ee
Here $c_n$ is the normalization of the twist operators, and the summation over $K$ is on all the independent quasiprimary operators $\Phi_K$ in $\CFT^n$, and there are definitions of coefficients
\be
a_K^r\equiv \f{C_{h_K+r-1}^r}{C_{2h_K+r-1}^r}, ~~ \bar a_K^s\equiv\f{C_{\bar h_K+s-1}^s}{C_{2\bar h_K+s-1}^s},
\ee
where the binomial coefficient is
\be
C_x^y=\f{\G(x+1)}{\G(y+1)\G(x-y+1)},
\ee
and the OPE coefficient $d_K$ can be determined by the one-point function on the Riemann surface $\mc R_{n,1}$ \cite{Calabrese:2010he}
\be \label{e3}
d_K=\f{1}{\a_K\ell^{h_K+\bar h_K}} \lim_{z\to\inf}z^{2 h_K}\bar z^{2\bar h_K}\lag \Phi_K(z,\bar z) \rag_{\mc R_{n,1}},
\ee
with $\a_K$ being the normalization coefficient of $\Phi_K$. The quasiprimary operators in CFT$^n$ can be constructed by the quasiprimary operators in each copy of CFT.

For the vacuum conformal family, to level 8 the $\CFT^n$ quasiprimary operators have been constructed in \cite{Chen:2013kpa,Chen:2013dxa}. In this paper we only need the ones to level 6 that have nonvanishing expectation values on the torus.
Because of the translation symmetry, one-point functions on torus are coordinate-independent, and so the derivatives of an operator have vanishing expectation values on the torus. This simplifies the discussion. To level 6, the holomorphic $\CFT^n$ quasiprimary operators that have nonvanishing expectation values on torus are listed as follows.
\begin{itemize}
  \item At level 2 we have $T_j$.
  \item At level 4 we have $\mA_j$, and $T_{j_1}T_{j_2}$ with $j_1 < j_2$.
  \item At level 6 we have $\mB_j$, $\mD_j$, $T_{j_1}\mA_{j_2}$ with $j_1 \neq j_2$, and $T_{j_1}T_{j_2}T_{j_3}$ with $j_1<j_2<j_3$.
\end{itemize}
Here the subscript $j$ labels the replica and  all the $j$'s take values from 0 to $n-1$. The detailed discussion on the quasiprimary operators can be found in Appendix~\ref{AppBasics}. Note that operators at level 5 and some of operators at level 6 have vanishing expectation values on the torus so that they can be ignored. The needed OPE coefficients $d_K$ have been calculated in \cite{Chen:2013kpa,Chen:2013dxa}
\bea
&& d_T=\frac{n^2-1}{12n^2},                                                        ~~
   d_\mA=\frac{(n^2-1)^2}{288 n^4},                                                ~~
   d_\mB=-\frac{(n^2-1)^2 \left(2 n^2(35 c+61)-93\right)}{10368 n^6(70 c+29)},     \nn\\
&& d_\mD=\frac{(n^2-1)^3}{10368 n^6},                                              ~~
   d_{TT}^{j_1j_2}=\f{1}{8n^4c}\f{1}{s^4_{j_1j_2}}+\f{(n^2-1)^2}{144n^4},          ~~
   d_{T\mA}^{j_1j_2}=\f{n^2-1}{96n^6c}\f{1}{s^4_{j_1j_2}}+\f{(n^2-1)^3}{3456n^6},  \\
&& d_{TTT}^{j_1j_2j_3}=-\f{1}{8n^6c^2}\f{1}{s^2_{j_1j_2}s^2_{j_2j_3}s^2_{j_3j_1}}
                       +\f{n^2-1}{96n^6c} \lt( \f{1}{s^4_{j_1j_2}}+\f{1}{s^4_{j_2j_3}}+\f{1}{s^4_{j_3j_1}} \rt) \nn
                       +\f{(n^2-1)^3}{1728n^6},
\eea
with
\be
s_{j_1j_2}=\sin\f{(j_1-j_2)\pi}{n}.
\ee
We may introduce
\bea
&& b_T=n d_T, ~~ b_\mA=n d_\mA, ~~ b_\mB=n d_\mB, ~~ b_\mD=n d_\mD,                            \nn\\
&& b_{TT}=\sum_{j_1<j_2} d_{TT}^{j_1j_2}, ~~
   b_{T\mA}=\sum_{j_1\neq j_2} d_{T\mA}^{j_1j_2}, ~~
   b_{TTT}=\sum_{j_1<j_2<j_3} d_{TTT}^{j_1j_2j_3},
\eea
and find
\bea
&& \hspace{-10mm}
   b_{TT}=\frac{(n^2-1) [5 c (n+1) (n-1)^2+2 (n^2+11)]}{1440 c n^3},                                ~~
   b_{T\mA}=\frac{(n^2-1)^2 [5 c (n+1) (n-1)^2+4 (n^2+11)]}{17280 c n^5},                           \nn\\
&& \hspace{-10mm}
   b_{TTT}=\frac{(n-2) (n^2-1) [35 c^2 (n+1)^2 (n-1)^3+42 c (n^2-1) (n^2+11)-16 (n+2) (n^2+47)]}
                {362880 c^2 n^5}.
\eea
It is similar for the quasiprimary operators in the antiholomorphic sector.



For other modules, the construction of quasiprimary operators is straightforward. In the following discussion, we mainly consider the primary operator $\phi$ with conformal weight $(h_\phi,0)$, in which case  $h_\phi$ has to be an integer or a half-integer. Such operators appear in the CFT with $\cal W$ symmetry or the superconformal symmetry. In the OPE of the twist operators, there are the quasiprimary operators from the module $\phi$,  among which the operators $\phi_{j_1}\phi_{j_2}$  with $j_1<j_2$ have the lowest scaling dimension. The normalization of $\phi$ is chosen to be $\a_\phi$. We have the OPE coefficient
\be
d_{\phi\phi}^{j_1j_2}=\f{\ii^{2h_\phi}}{\a_\phi(2n)^{2h_\phi}}\f{1}{s_{j_1j_2}^{2h_\phi}}.
\ee
For later convenience, we introduce
\be
b_{\phi\phi} = \sum_{j_1<j_2} d_{\phi\phi}^{j_1j_2}.
\ee
Similarly, we have $d_{\bar\phi\bar\phi}^{j_1j_2}$ and $b_{\bar\phi\bar\phi}$ for the antiholomorphic sector.

\subsection{R\'enyi entropy}

We consider the R\'enyi entropy of one short interval $A=[0,\ell]$ on a torus $\mT$ with the complex coordinate $z$.
The modulus of the torus is $\t=\t_1+\ii\t_2$, with $\t_1$ and $\t_2$ being real. The modulus $\t$ is related to the inverse temperature $\b$ and chemical potential $\m_E$ by
\be
\t_1=\f{\b\m_E}{L}, ~~ \t_2=\f{\b}{L}.
\ee
Note that here we are working in Euclidean space, and the Euclidean chemical potential $\m_E$ is real.
If we want the Minkowski result, in the final R\'enyi entropy we have to make the analytical continuation $\m_E=-\ii\m$ with the Minkowski chemical potential $\m$ being real.
The spatial period of the torus is $L$ so  we have
\be
z \simeq z+ L(r +s\t), ~~ r,s\in\Z.
\ee
Without loss of generality,  we may set $L=1$.
As usual we define
\be q \equiv\ep^{2\pi\ii\t}.
\ee

The partition function is given by
\be
\tr_A\r_A^n = \langle \s(\ell,\ell)\td \s(0,0)\rangle_\mT
            = \f{c_n}{\ell^{2(h_\s+\bar h_\s)}} \sum_K d_K \ell^{h_K+\bar h_K} \lag \Phi_K(0,0) \rag_\mT.
\ee
Due to OPE of the twist operators, we have summation over the one-point functions of the quasiprimary operators in CFT$^n$. In principle, this could be applied to any CFT. Here we are more interested in the large $c$ CFT dual to
the AdS$_3$ gravity. In this case, the contribution is dominated by the vacuum module. To the level we are interested in, all quasiprimary operators in the OPE are the products of the quasiprimary operators in a single CFT on different replica. Therefore, their expectation values are just the products of the ones on the torus. The resulting partition function is
\bea
&& \tr_A\r_A^n=\f{c_n}{\ell^{2(h_\s+\bar h_\s)}}
             \bigg\{ 1+b_T\lag T\rag_\mT \ell^2
                    + \Big(b_\mA\lag\mA\rag_\mT+b_{TT}\lag T\rag_\mT^2 \Big) \ell^4
                    + \Big(b_\mB\lag\mB\rag_\mT + b_\mD\lag\mD\rag_\mT + b_{T\mA}\lag T\rag_\mT\lag\mA\rag_\mT   \nn\\
&&\phantom{\tr_A\r_A^ n=\f{c_n}{\ell^{2(h_\s+\bar h_\s)}}~}
                    +b_{TTT}\lag T\rag_\mT^3\Big) \ell^6 +O(\ell^8)
                    + \sum_\phi \Big[ \ell^{2h_\phi} \Big(b_{\phi\phi}\lag\phi\rag_\mT^2 +O(\ell^2) \Big)+O(\ell^{3h_\phi}) \Big] \bigg\}     \\
&&\phantom{\tr_A\r_A^n=\f{c_n}{\ell^{2(h_\s+\bar h_\s)}}}\hspace{-4mm}\times
            \bigg\{ 1+b_T\lag \bar T\rag_\mT \ell^2
                   + \Big(b_\mA\lag\bar\mA\rag_\mT+b_{TT}\lag \bar T\rag_\mT^2 \Big) \ell^4
                   + \Big(b_\mB\lag\bar\mB\rag_\mT + b_\mD\lag\bar\mD\rag_\mT
                   + b_{T\mA}\lag \bar T\rag_\mT\lag\bar\mA\rag_\mT                                              \nn\\
&&\phantom{\tr_A\r_A^ n=\f{c_n}{\ell^{2(h_\s+\bar h_\s)}}~}
                   + b_{TTT}\lag \bar T\rag_\mT^3\Big) \ell^6 + O(\ell^8)
                   + \sum_{\bar\phi} \Big[ \ell^{2h_\phi} \Big(b_{\bar\phi\bar\phi}\lag\bar\phi\rag_\mT^2 +O(\ell^2) \Big)+O(\ell^{3h_\phi}) \Big] \bigg\}, \nn
\eea
with the two summations being over all the nonidentity holomorphic and antiholomorphic primary operators respectively. Consequently
we can get the short interval expansion of the single interval R\'enyi entropy
\bea \label{Snexpansion}
&&\hspace{-6mm}
S_n= \f{c(n+1)}{6n}\log\f{\ell}{\e}   -\f{1}{n-1} \bigg\{ b_T\big(\lag T \rag_\mT+\lag \bar T \rag_\mT\big)\ell^2
       +\Big[  b_\mA \big(\lag\mA\rag_\mT+\lag\bar\mA\rag_\mT\big)
              + \big(b_{TT}-\f12 b_T^2\big)\big(\lag T\rag_\mT^2+\lag\bar T\rag_\mT^2\big) \Big]\ell^4                                \nn\\
&&\hspace{-6mm}\phantom{S_n=}
    +\Big[  b_\mB \big(\lag\mB\rag_\mT+\lag\bar\mB\rag_\mT\big)+ b_\mD \big(\lag\mD\rag_\mT+\lag\bar\mD\rag_\mT\big)
          +(b_{T\mA}-b_T b_\mA)\big(\lag T\rag_\mT \lag\mA\rag_\mT+\lag\bar T\rag_\mT\lag \bar\mA\rag_\mT\big)                        \\
&&\hspace{-6mm}
\phantom{S_n=}
          + \big(b_{TTT}-b_T b_{TT}+\f13 b_T^3\big)\big(\lag T\rag_\mT^3+\lag\bar T\rag_\mT^3\big)\Big]\ell^6
+ O(\ell^8) + \sum_{\phi/\bar\phi} \ell^{2h_\phi}\Big[  b_{\phi\phi}\lag\phi\rag_\mT^2
                                                      + b_{\bar\phi\bar\phi}\lag\bar\phi\rag_\mT^2
                                                      + O(\ell^2)
                                                      + O(\ell^{h_\phi}) \Big] \bigg\}. \nn
\eea
Note that $\phi$ and $\bar\phi$ appear in pairs, and the summation above is over such pairs.
Therefore, the computation of the single interval R\'enyi entropy is reduced to the  one-point functions on the torus.  Note that because of  translation symmetry the one-point functions on torus are independent of the coordinate.

On a torus, the one-point function of the stress tensor is given by\cite{Eguchi:1986sb}
\be \label{T}
\lag T \rag_\mT=2\pi\ii\p_\t\log Z,
\ee
with $Z$ being the partition function on the torus.
From the Ward identity on torus \cite{Eguchi:1986sb}, we get
\bea \label{TTnTTT}
&& \lag T(y)T(z) \rag_\mT=\f{c}{12}\wp''(y-z) + \big( \lag T \rag_\mT+ 2\wp(y-z)+4\eta_1+2\pi\ii\p_\t \big)\lag T \rag_\mT, \nn\\
&& \lag T(x)T(y)T(z) \rag_\mT = \f{c}{12}(\wp''(x-y)+\wp''(x-z))\lag T \rag_\mT                                             \\
&& \phantom{\lag T(x)T(y)T(z) \rag_\mT =}
   +\big( \lag T \rag_\mT+ 2\wp(x-y)+2\wp(x-z)+8\eta_1+2\pi\ii\p_\t \big)\lag T(y)T(z) \rag_\mT                                      \nn\\
&& \phantom{\lag T(x)T(y)T(z) \rag_\mT =}
   +\big( \z(x-y) - \z(x-z) + 2\eta_1 y -2\eta_1 z \big) \lag \p T(y)T(z) \rag_\mT,                                            \nn
\eea
with $\wp(x)$ being the Weierstrass P function, $\z(x)$ being the Weierstrass $\z$ function, and
\be \label{eta1}
\eta_1=4\pi^2 \Big( \f{1}{24}-\sum_{k=1}^\inf\f{k q^k}{1-q^k} \Big).
\ee
From (\ref{TTnTTT}) and the definitions in (\ref{Adef}) and (\ref{BDdef}), we get
\bea \label{ABD}
&& \lag \mA \rag_\mT = \frac{c g_2}{120} + ( \lag T\rag_\mT +4\eta_1 + 2\pi\ii\p_\t) \lag T\rag_\mT,                          ~~
   \lag \mB \rag_\mT = -\frac{9c g_3}{70}-\frac{9g_2}{25} \lag T \rag_\mT,                                                    \nn\\
&& \lag \mD \rag_\mT = -\frac{3 c (5 c+22) g_3}{5 (70 c+29)} +\frac{c g_2\eta_1}{15} +\frac{c}{60} \pi\ii\p_\t g_2
                       +2\pi \big[ 3(\lag T \rag_\mT+ 4 \eta_1)\ii\p_\tau -2\pi \p_\tau^2\big]\lag T \rag_\mT                   \\
&& \phantom{\lag \mD \rag_\mT =}
                       +\frac{(42c^2-61c-836) g_2}{24 (70 c+29)} \lag T\rag_\mT
                       +\big[(\lag T \rag_\mT+4\eta_1)(\lag T \rag_\mT+8\eta_1)+8\pi\ii\p_\tau\eta_1 \big] \lag T \rag_\mT.   \nn
\eea
Here we have introduced
\bea \label{g2g3}
&& g_2=\f{4\pi^4}{3}\Big( 1+240\sum_{k=1}^\inf\s_3(k)q^k \Big),         \nn\\
&& g_3=\f{8\pi^6}{27}\Big( 1-504\sum_{k=1}^\inf\s_5(k)q^k \Big),
\eea
where $\s_a(k)$ is the divisor function.

Note that the short interval expansion (\ref{Snexpansion}) holds for any value of torus modulus $\t$, but the expansions in (\ref{eta1}) and (\ref{g2g3}) may not converge. For any $\t$ we can always find an $\SL(2,\Z)$ modular transformation\footnote{One should not confuse the integer $c$ here with the central charge.}
\be
\t'=\f{a\t+b}{c\t+d}, ~~ q'=\ep^{2\pi\t'},
\ee
with $a,b,c,d\in\Z$ and $ad-bc=1$, and make the expansions in (\ref{eta1}) and (\ref{g2g3}) be well-defined. The transformation rules are
\bea \label{rule1}
&& \eta_1(\t)=\f{\eta_1(\t')}{(c\t+d)^2}+\f{c\pi\ii}{c\t+d}, \nn\\
&& g_2(\t)=\f{g_2(\t')}{(c\t+d)^4}, ~~
   g_3(\t)=\f{g_3(\t')}{(c\t+d)^6}.
\eea
For consistency of the CFT, the partition function $Z$ is invariant under the $\SL(2,\Z)$ transformation
\be \label{rule2}
Z(\t)=Z(\t').
\ee

 At a low temperature, the holomorphic sector of the density matrix of the vacuum conformal family can be expanded as
\be \label{rvac}
\r_\vac = |0\rag \lag0| + \f{q^2}{\a_T}|T\rag\lag T\rag
         +\f{q^3}{\a_{\p T}}|\p T\rag\lag \p T\rag
         +\f{q^4}{\a_{\p^2T}}|\p^2T\rag\lag\p^2T\rag
         +\f{q^4}{\a_{\mA}}|\mA\rag\lag\mA\rag
         +O(q^5).
\ee
More generally, there could be other modules in the theory. Thus one can add other holomorphic conformal families and have the density matrix
\be
\r=\r_\vac+\sum_\phi\r_\phi,
\ee
with the definition
\be
\r_\phi=q^{h_\phi}\sum_{k=0}^\inf q^k \sum_{i,j}G_{\phi,k}^{ij} |\phi,k,i\rag\lag\phi,k,j|.
\ee
Here $h_\phi$ is the conformal weight, $k$ denotes the levels in the $\phi$ conformal family, and $G_{\phi,k}^{ij}$ is the inverse of the matrix $\lag\phi,k,i|\phi,k,j\rag$. The one-point function of an operator on the torus is just the thermal expectation value of the operator on the cylinder, which can be calculated by mapping the cylinder to a complex plane. In other words, one can cut open the torus and insert a complete set of state basis at the cut such that the one-point function on the torus reduces to a summation of three-point functions. In Neveu-Schwarz (NS) sector, a holomorphic operator with half-integer conformal weight has vanishing one-point function. For a nonidentity holomorphic primary operator $\phi$ with integer conformal weight $h_\phi$, one can get the one-point function on the torus
\be \label{e34}
\lag\phi\rag_\mT = \f{(2\pi\ii)^{h_\phi}}{\tr\r}\sum_\psi q^{h_\psi}\sum_{k=0}^\inf q^k \sum_{i,j}G_{\psi,k}^{ij} \lag\psi,k,i|\phi_0|\psi,k,j\rag,
\ee
with $\phi_0$ being the $0$-th mode of $\phi$. In the low temperature expansion, the leading order contribution of the one-point function is
\be
\lag\phi\rag_\mT = (2\pi\ii)^{h_\phi} q^{h_\psi}\f{C_{\psi\phi\psi}}{\a_\psi}+\cdots,
\ee
with $\psi$ being the primary operator that has the lowest conformal weight $h_\psi$ among the ones with nonvanishing structure constant $C_{\psi\phi\psi}$ and $\a_\psi$ being its normalization. In the R\'enyi entropy (\ref{Snexpansion}) we have the following term in the expansion of $\ell$ and $q$
\be \label{AddTerm}
\ell^{2h_\phi} b_{\phi\phi} \lag\phi\rag_\mT^2  \sim  \ell^{2h_\phi} q^{2h_\psi} \f{(C_{\psi\phi\psi})^2}{\a_\phi\a_\psi^2} + \cdots.
\ee
The behavior of the term in the expansion of $1/c$ has to be discussed case by case.

\section{Vacuum conformal family}\label{SecVac}

In the discussion above, the prescription can be applied to any CFT at any temperature.
In this section we focus on the contributions of the vacuum conformal family to the R\'enyi entropy, and we temporarily ignore the contributions from other conformal families. We would like to compute the entropy at both the low and high temperature limits, in order to compare with the results got in \cite{Chen:2014unl,Chen:2015uia}.

\subsection{Low temperature limit}

 At a low temperature we first consider the case without chemical potential, and we have\footnote{Note that the length of the spatial direction $L$ has been set to be 1.}
\be
q=\bar q=\ep^{-2\pi\b} \ll 1.
\ee
We will briefly discuss the case with chemical potential in the end of this subsection.
On the gravity side the R\'enyi entropy can be calculated to the order $q^4$ \cite{Barrella:2013wja,Chen:2014unl,Chen:2015uia}, and the gravity result has been confirmed from the CFT side \cite{Cardy:2014jwa,Chen:2014unl,Chen:2015uia}.
We may expand the holographic R\'enyi entropy by the length of the interval $\ell$.
The classical part is
\bea \label{Sncl}
&& \hspace{-10mm}
   S_n^\cl = \frac{c (n+1)}{6 n} \log\f{\ell}{\e}  -\frac{\pi^2 c (n+1)}{36 n}\ell^2
                    + \Big( -\frac{\pi^4 c (n+1)}{1080 n}
                            -\frac{\pi^4 c (n-1) (n+1)^2 }{9 n^3}q^2
                            -\frac{4\pi^4 c (n-1) (n+1)^2}{9 n^3}q^3                                              \nn\\
&& \hspace{-10mm} \phantom{S_n^\cl =}
                            -\frac{11\pi^4 c (n-1) (n+1)^2}{9 n^3}q^4+O(q^5) \Big)\ell^4
                   + \Big( -\frac{\pi^6 c (n+1)}{17010 n}
                           +\frac{2\pi^6 c (n-1) (n+1)^2}{27 n^3}q^2                                              \\
&& \hspace{-10mm} \phantom{S_n^\cl =}
                           +\frac{20\pi^6 c (n-1) (n+1)^2}{27 n^3}q^3
                           +\frac{2\pi^6 c (n-1) (n+1)^2 (48 n^2-1)}{27 n^5} q^4+O(q^5) \Big)\ell^6 + O(\ell^8),  \nn
\eea
and the one-loop part is
\bea \label{Sn2oloop}
&& \hspace{-10mm}
    S_{n,(2)}^\oloop=  \Big( \frac{4\pi^2 (n+1)}{3 n}q^2
                     +\frac{2\pi^2 (n+1)}{n}q^3
                     +\frac{4\pi^2 (n+1)}{n}q^4 + O(q^5) \Big)\ell^2
               + \Big( -\frac{2\pi^4 (n+1) (9 n^2-11)}{45 n^3}q^2                                \nn\\
&& \hspace{-10mm} \phantom{S_{n,(2)}^\oloop=}
                       -\frac{2\pi^4 (n+1) (41 n^2-44)}{45 n^3}q^3
                       -\frac{2\pi^4 (n+1) (49 n^2-51)}{15 n^3}q^4 + O(q^5) \Big) \ell^4        \\
&& \hspace{-10mm} \phantom{S_{n,(2)}^\oloop=}
               + \Big(  \frac{4\pi^6 (n+1) (17 n^4-46 n^2+31)}{945 n^5}q^2
                       +\frac{2\pi^6 (n+1) (492 n^4-1013 n^2+527)}{945 n^5}q^3                   \nn\\
&& \hspace{-10mm} \phantom{S_{n,(2)}^\oloop=}
                       +\frac{4\pi^6 (n+1) (1654 n^4-2903 n^2+1255)}{945 n^5}q^4 + O(q^5) \Big)\ell^6 + O(\ell^8). \nn
\eea
The calculation of the R\'enyi entropy to higher orders of $q$ is very cumbersome. However, since it is exact in $\ell$, one can get the R\'enyi entropy to the higher order of $\ell$ very easily.

On the CFT side, if we only consider the vacuum conformal family, we have the partition function
\be \label{Z}
Z=q^{-\f{c}{24}}\bar q^{-\f{c}{24}}\prod_{k=2}^\inf\f{1}{(1-q^k)(1-\bar q^k)}.
\ee
From (\ref{T}) and (\ref{ABD}) we get the one-point functions\footnote{The low temperature expansion of these one-point functions can be reproduced in a different method, as shown in Appendix~\ref{AppLow}.}
\bea \label{onepointfunctions}
&&\hspace{-10mm}
   \lag T   \rag_\mT = \frac{\pi^2 c}{6}-8 \pi^2 q^2-12 \pi^2 q^3-24 \pi^2 q^4+O(q^5),  \nn\\
&&\hspace{-10mm}
    \lag \mA \rag_\mT = \frac{\pi^4 c (5 c+22)}{180}+\frac{8\pi^4(5c+22)}{3} q^2+12 \pi^4 (5 c+22) q^3+168 \pi^4 (c+6) q^4+O(q^5) ,\nn\\
&&\hspace{-10mm}
    \lag \mB \rag_\mT = -\frac{62\pi^6 c}{525} +\frac{96\pi^6(120c+1) }{25} q^2
                       +\frac{144\pi^6 (720 c+161)}{25} q^3+\frac{288\pi^6 (1640 c+841)}{25} q^4+O(q^5) ,\\
&&\hspace{-10mm}
   \lag \mD \rag_\mT = \frac{\pi^6 c (2 c-1) (5 c+22) (7 c+68)}{216 (70 c+29)}
                      +\frac{22 \pi^6 (2 c-1) (5 c+22) (7 c+68)}{3 (70 c+29)}q^2 \nn\\
&&\hspace{-10mm}\phantom{\lag \mD \rag_\mT =}
                      +\frac{31 \pi^6 (2 c-1) (5 c+22) (7 c+68)}{70 c+29}q^3
                      +\frac{2 \pi^6 (2 c-1) (7 c+68) (215 c-638)}{70 c+29}q^4+O(q^5). \nn
\eea
There are similar results for the antiholomorphic operators. Using (\ref{Snexpansion}) and doing large $c$ expansion, we obtain the R\'enyi entropy, which could be organized by the powers of $1/c$
\be
S_{n}^{(2)}=S_{n}^\cL+S_{n,(2)}^{\NL}+S_{n,(2)}^{\NNL}+\cdots.
\ee
Here the notation $(2)$ denotes that we are considering the large $c$ CFT dual to the pure AdS$_3$ gravity, and the superscripts $\cL$, $\NL$, $\NNL$ denote the leading order, next-to-leading order and next-to-next-to-leading order respectively.
We did not have subscript $(2)$ in the leading part, since it is universal and does not depend on the operator content of the CFT.
The leading contribution $S_{n}^\cL$ is proportional to $c$, and it is in perfect agreement with  the classical part of the holographic R\'enyi entropy $S_n^\cl$ in (\ref{Sncl}). The next-to-leading part $S_{n,(2)}^{\NL}$ is of order one, and it agrees exactly with the one-loop gravitational part $S_{n,(2)}^\oloop$ in (\ref{Sn2oloop}). The next-to-next-to-leading part $S_{n,(2)}^{\NNL}$ is proportional to $1/c$
\be \label{Sn2NNL}
S_{n,(2)}^{\NNL} =  \Big( -\frac{8\pi^4(n+1) (n^2+11)}{45n^3c}q^4 + O(q^5) \Big)\ell^4
                   +\Big( \frac{8\pi^6(n+1)(26 n^4+271 n^2-345)}{945n^5c}q^4 + O(q^5) \Big)\ell^6
                   +O(\ell^8),
\ee
which is expected to match the two-loop correction to the holographic R\'enyi entropy.
Note that expansion of R\'enyi entropy to the higher orders of $q$ is straightforward and can be done easily. However, the expansion of R\'enyi entropy to higher orders of $\ell$ is cumbersome. The parameter region of the method in this paper is different from that of the gravity and CFT methods in \cite{Barrella:2013wja,Cardy:2014jwa,Chen:2014unl,Chen:2015uia}.


When the chemical potential is turned on, one need to make the replacement
\be
q^m \to \f{q^m+\bar q^m}{2} = \cos (2\pi m\b\m_E)\ep^{-2\pi m\b}, ~~ m\in \Z
\ee
in the R\'enyi entropy. In the Minkowski spacetime, we have to make the analytical continuation $\m_E = -\ii\m$, and the replacement becomes
\be
q^m \to \f{q^m+\bar q^m}{2} = \cosh (2\pi m\b\m)\ep^{-2\pi m\b}, ~~ m\in \Z.
\ee
The CFT at a low temperature with a chemical potential is dual to the thermal AdS with angular momentum. It would be nice to see if the R\'enyi entropy can be reproduced on the gravity side.

\subsection{High temperature limit without chemical potential}

At the high temperature limit without chemical potential, the classical and one-loop gravity results are just making the following replacement\cite{Barrella:2013wja} in (\ref{Sncl}) and (\ref{Sn2oloop})
\be
\ell \to \f{\ell}{\ii\b}, \hs{5ex} q=\ep^{2\pi\ii\t}=\ep^{-2\pi\b} \to q'=\ep^{2\pi\ii\t'}=\ep^{-2\pi/\b},
\ee
except the $\ell$ in the logarithmic function.
The classical part becomes
\bea\label{SnclHT}
&& \hspace{-10mm}
    S_n^\cl = \frac{c (n+1)}{6 n} \log\f{\ell}{\e}  +\frac{\pi^2 c (n+1)}{36 n}\f{\ell^2}{\b^2}
                    + \Big( -\frac{\pi^4 c (n+1)}{1080 n}
                            -\frac{\pi^4 c (n-1) (n+1)^2 }{9 n^3}q'^2
                            -\frac{4\pi^4 c (n-1) (n+1)^2}{9 n^3}q'^3                                              \nn\\
&& \hspace{-10mm} \phantom{S_n^\cl =}
                            -\frac{11\pi^4 c (n-1) (n+1)^2}{9 n^3}q'^4+O(q'^5) \Big)\f{\ell^4}{\b^4}
                   - \Big( -\frac{\pi^6 c (n+1)}{17010 n}
                           +\frac{2\pi^6 c (n-1) (n+1)^2}{27 n^3}q'^2                                              \\
&& \hspace{-10mm} \phantom{S_n^\cl =}
                           +\frac{20\pi^6 c (n-1) (n+1)^2}{27 n^3}q'^3
                           +\frac{2\pi^6 c (n-1) (n+1)^2 (48 n^2-1)}{27 n^5} q'^4+O(q'^5) \Big)\f{\ell^6}{\b^6} + O(\ell^8),  \nn
\eea
and the one-loop part becomes
\bea \label{Sn2oloopHT}
&& \hspace{-10mm}
   S_{n,(2)}^\oloop=  -\Big( \frac{4\pi^2 (n+1)}{3 n}q'^2
                     +\frac{2\pi^2 (n+1)}{n}q'^3
                     +\frac{4\pi^2 (n+1)}{n}q'^4 + O(q'^5) \Big)\f{\ell^2}{\b^2}
               + \Big( -\frac{2\pi^4 (n+1) (9 n^2-11)}{45 n^3}q'^2                                \nn\\
&& \hspace{-10mm} \phantom{S_{n,(2)}^\oloop=}
                       -\frac{2\pi^4 (n+1) (41 n^2-44)}{45 n^3}q'^3
                       -\frac{2\pi^4 (n+1) (49 n^2-51)}{15 n^3}q'^4 + O(q'^5) \Big) \f{\ell^4}{\b^4}        \\
&& \hspace{-10mm} \phantom{S_{n,(2)}^\oloop=}
               - \Big(  \frac{4\pi^6 (n+1) (17 n^4-46 n^2+31)}{945 n^5}q'^2
                       +\frac{2\pi^6 (n+1) (492 n^4-1013 n^2+527)}{945 n^5}q'^3                   \nn\\
&& \hspace{-10mm} \phantom{S_{n,(2)}^\oloop=}
                       +\frac{4\pi^6 (n+1) (1654 n^4-2903 n^2+1255)}{945 n^5}q'^4 + O(q'^5) \Big)\f{\ell^6}{\b^6} + O(\ell^8). \nn
\eea

On the CFT side we make the modular transformation
\be
\t'=-\f1\t,
\ee
and expand the R\'enyi entropy in $q'$. Using the transformation rules (\ref{rule1}) and (\ref{rule2}), we get the high temperature expansion of the one-point functions (\ref{T}) and (\ref{ABD})
\bea \label{onepointfunctionsHT}
&& \hspace{-10mm}
   \lag T \rag_\mT = -\frac{\pi^2 c}{6 \beta^2}
                     +\frac{8 \pi^2}{\beta^2}q'^2
                     +\frac{12 \pi^2}{\beta^2}q'^3
                     +\frac{24 \pi^2}{\beta^2}q'^4
                     +O(q'^5),  \nn\\
&& \hspace{-10mm}
   \lag \mA \rag_\mT = \frac{\pi^4 c (5 c+22)}{180 \beta^4}
                      +\frac{8 \pi^4 (5 c+22)}{3 \beta^4}q'^2
                      +\frac{12 \pi^4 (5 c+22)}{\beta^4}q'^3
                      +\frac{168 \pi^4 (c+6)}{\beta^4}q'^4
                      +O(q'^5) ,  \nn\\
&& \hspace{-10mm}
   \lag \mB \rag_\mT = \frac{62 \pi^6 c}{525 \beta^6}
                      -\frac{96\pi^6 (120 c+1)}{25 \beta^6}q'^2
                      -\frac{144 \pi^6 (720 c+161)}{25 \beta^6}q'^3
                      -\frac{288 \pi^6 (1640 c+841)}{25 \beta^6}q'^4
                      +O(q'^5),  \\
&& \hspace{-10mm}
   \lag \mD \rag_\mT = -\frac{\pi^6 c (2 c-1) (5 c+22) (7 c+68)}{216(70 c+29)\beta^6}
                       -\frac{22 \pi^6 (2 c-1) (5 c+22) (7 c+68)}{3(70 c+29)\beta^6}q'^2 \nn\\
&& \hspace{-10mm}\phantom{\lag \mD \rag_\mT =}
                       -\frac{31 \pi^6 (2 c-1) (5 c+22) (7 c+68)}{(70 c+29)\beta^6}q'^3
                       -\frac{2 \pi^6 (2 c-1) (7 c+68) (215 c-638)}{(70 c+29)\beta^6}q'^4
                       +O(q'^5).  \nn
\eea
Note that if we restore $L$ in (\ref{onepointfunctions}), then Eq.(\ref{onepointfunctionsHT}) equal Eq. (\ref{onepointfunctions}) after the simple substitutions
\be
L\to\ii\b, \hs{3ex} q \to q'.
\ee
There are similar results for the antiholomorphic operators. Using (\ref{Snexpansion}) and organizing the R\'enyi entropy by the powers of $1/c$, we find
\be \label{e32}
S_{n}^{(2)}= S_{n}^\cL+S_{n,(2)}^{\NL}+S_{n,(2)}^{\NNL}+\cdots.
\ee
Now $S_{n}^\cL$ and $S_{n,(2)}^{\NL}$ are in exact match with  the classical part $S_n^\cl$ (\ref{SnclHT}) and the one-loop part $S_{n,(2)}^\oloop$ (\ref{Sn2oloopHT}) respectively. The next-to-next-to-leading part is
\be
S_{n,(2)}^\NNL = \Big( -\frac{8\pi^4(n+1) (n^2+11)}{45n^3c}q'^4 + O(q'^5) \Big) \f{\ell^4}{\b^4}
                -\Big( \frac{8\pi^6(n+1)(26 n^4+271 n^2-345)}{945n^5c}q'^4 + O(q'^5) \Big) \f{\ell^6}{\b^6}
                +O(\ell^8).
\ee
The CFT at a high temperature with vanishing chemical potential is dual to nonrotating BTZ (Ba\~nados-Teitelboim-Zanelli) black hole \cite{Banados:1992wn}.

\subsection{High temperature limit with chemical potential}

When we turn the chemical potential on, the calculation is similar. Firstly, we have
\bea
&& \t=\ii\b_L, ~~ \b_L=\b(1-\ii\m_E)=\b(1-\m),       \\
&& \bar\t=-\ii\b_R, ~~ \b_R=\b(1+\ii\m_E)=\b(1+\m).  \nn
\eea
Note that in the Euclidean space the left- and right-moving inverse temperatures $\b_{L,R}$ are complex, but they are real in the Minkowski spacetime. One can see this point, for example, in \cite{Castro:2014tta}. At a high temperature $\b$ is small, and we work in the parameter region that $\m$ is of order one. We make the modular transformation
\bea
&& \t'=-\f{1}{\t}=\f{\ii}{\b_L}, ~~
   q'=\ep^{2\pi\ii\t'}=\ep^{-2\pi/\b_L},            \\
&& \bar\t'=-\f{1}{\bar\t}=-\f{\ii}{\b_R}, ~~
   \bar q'=\ep^{-2\pi\ii\bar\t'}=\ep^{-2\pi/\b_R}.  \nn
\eea
The one-point functions $\lag T \rag_\mT$, $\lag \mA \rag_\mT$, $\lag \mB \rag_\mT$, $\lag \mD \rag_\mT$ are formally the same as (\ref{onepointfunctionsHT}) by setting $\b \to \b_L$.
The one-point functions of the antiholomorphic operators $\lag \bar T \rag_\mT$, $\lag \bar\mA \rag_\mT$, $\lag \bar\mB \rag_\mT$, $\lag \bar\mD \rag_\mT$ can be got from $\lag T \rag_\mT$, $\lag \mA \rag_\mT$, $\lag \mB \rag_\mT$, $\lag \mD \rag_\mT$ by setting
$\b_L \to \b_R$, $q' \to \bar q'$.
We get the R\'enyi entropy by the powers of $1/c$
\be \label{e33}
S_{n}^{(2)}= S_{n}^{\cL,L} + S_{n}^{\cL,R}
          +S_{n,(2)}^{\NL,L} + S_{n,(2)}^{\NL,R}
          +S_{n,(2)}^{\NNL,L} + S_{n,(2)}^{\NNL,R} + \cdots.
\ee
Here the superscripts $L,R$ denote the R\'enyi entropy from the holomorphic and antiholomorphic sectors respectively, or, in terms of the Minkowski CFT, the left- and right-moving sectors. Formally, $S_{n}^{\cL,L}$, $S_{n,(2)}^{\NL,L}$ and $S_{n,(2)}^{\NNL,L}$ are just half of the previous subsection results $S_{n}^{\cL}$, $S_{n,(2)}^{\NL}$, and $S_{n,(2)}^{\NNL}$ in (\ref{e32}) after setting $\b \to \b_L$. Also $S_{n}^{\cL,R}$, $S_{n,(2)}^{\NL,R}$ and $S_{n,(2)}^{\NNL,R}$ are just $S_{n}^{\cL,L}$, $S_{n,(2)}^{\NL,L}$ and $S_{n,(2)}^{\NNL,L}$ after setting $\b_L\to\b_R$, $q'\to\bar q'$.

The CFT at the high temperature with a chemical potential is dual to the gravitational configuration of rotating BTZ black hole.
The holographic computation of the R\'enyi entropy in this case has not been worked out. However, the classical part of the entanglement entropy can be read from the length of a geodesic in the background of the BTZ black hole \cite{Hubeny:2007xt}
\be
S_\cl=\f{c}{6}\log\bigg( \f{\b_L\b_R}{\pi^2\e^2} \sinh\f{\pi\ell}{\b_L} \sinh\f{\pi\ell}{\b_R} \bigg).
\ee
Expanding the holographic entanglement entropy by small $\ell$, we get
\be
S_\cl = \f{c}{6} \bigg[  2 \log\f{\ell}{\e}
                       + \f{\pi^2\ell^2}{6} \bigg( \frac{1}{\beta_L^2}+\frac{1}{\beta_R^2} \bigg)
                       - \f{\pi^4\ell^4}{180} \bigg( \frac{1}{\beta_L^4}+\frac{1}{\beta_R^4} \bigg)
                       + O(\ell^6) \bigg].
\ee
This agrees with the above leading CFT result in (\ref{e33})
\be
S_\cl = \lim_{n\to 1} \Big( S_{n}^{\cL,L} + S_{n}^{\cL,R} \Big).
\ee

\section{CFT with $\mW\textbf{(2,3)}$ symmetry}\label{SecW23}

In the AdS$_3$/CFT$_2$ correspondence, we may extend the pure AdS$_3$ gravity to include the higher spin fields. With the higher spin fields, it turns out that after imposing appropriate asymptotical boundary conditions, the higher spin AdS$_3$ gravity could be dual to a 2D CFT with $\cal W$ symmetry \cite{Henneaux:2010xg,Campoleoni:2010zq}. In  the CFT with ${\cal W}$ symmetry, we have to consider the contributions from the operators $W$ and $\bar W$, besides the stress tensor $T$, $\bar T$. Here we only consider the CFT with $\mW(2,3)$ symmetry. The low temperature expansion of the additional R\'enyi entropy from the operator $W$  has been computed on both the gravity and CFT sides \cite{Chen:2015uia}. As we consider only the thermal state without higher spin chemical potential and correspondingly the gravitational configuration without higher spin hair, the contributions from the $W$ operators appear first in the next-to-leading order on the CFT side and in the one-loop correction on the bulk side. We may further expand such contributions from $W,\bar W$ fields by the length of the interval
 \bea \label{Sn3oloop}
&& S_{n,(3)}^\oloop= \Big( \frac{2 \pi^2 (n+1)}{n}q^3
                           +\frac{8\pi^2 (n+1)}{3 n}q^4
                           +O(q^5) \Big)\ell^2
                    +\Big( -\frac{2\pi^4 (n+1) (7 n^2-8)}{15 n^3}q^3                \nn\\
&& \phantom{S_{n,(3)}^\oloop=}
                           -\frac{8\pi^4 (n+1) (17 n^2-18)}{45 n^3}q^4
                           +O(q^5) \Big)\ell^4
                    +\Big( \frac{2\pi^6 (n+1) (128 n^4-313 n^2+191)}{945 n^5}q^3   \\
&& \phantom{S_{n,(3)}^\oloop=}
                          +\frac{4\pi^6 (n+1) (538 n^4-1107 n^2+573)}{945 n^5}q^4
                          +O(q^5) \Big)\ell^6
                   +O(\ell^8).                                                      \nn
\eea

On the CFT side, the torus partition function receives the contribution from $W$ operators as well
\be
Z=q^{-\f{c}{24}}\bar q^{-\f{c}{24}}\prod_{k=0}^\inf\f{1}{(1-q^{k+2})(1-q^{k+3})(1-\bar q^{k+2})(1-\bar q^{k+3})}.
\ee
Consequently, from (\ref{T}) and (\ref{ABD}) we read the one-point functions
\bea
&& \hspace{-7mm}
   \lag T \rag_\mT = \frac{\pi^2 c}{6}-8 \pi^2 q^2-24 \pi^2 q^3-40 \pi^2 q^4+O(q^5),\nn\\
&& \hspace{-7mm}
   \lag \mA \rag_\mT = \frac{\pi^4 c (5 c+22)}{180}
                      +\frac{8\pi^4 (5 c+22)}{3} q^2
                      +8 \pi^4 (7 c+50) q^3
                      +\frac{8\pi^4 (61 c+542)}{3} q^4
                      +O(q^5),\nn\\
&& \hspace{-7mm}
   \lag \mB \rag_\mT = -\frac{62\pi^6 c}{525}
                       +\frac{96\pi^6 (120 c+1)}{25} q^2
                       +\frac{2592\pi^6 (40 c+9)}{25} q^3
                       +\frac{96\pi^6 (984 c+577)}{5} q^4
                       +O(q^5) ,\\
&& \hspace{-7mm}
   \lag \mD \rag_\mT = \frac{\pi^6 c (2 c-1) (5 c+22) (7 c+68)}{216 (70 c+29)}
                      +\frac{22 \pi^6 (2 c-1) (5 c+22) (7 c+68)}{3(70 c+29)}q^2  \nn\\
&& \hspace{-7mm}\phantom{\lag \mD \rag_\mT =}
                      +\frac{6 \pi^6 (350 c^3+5661c^2-3702c-14648)}{70 c+29}q^3
                      +\frac{2 \pi^6 (8890 c^3+78695 c^2-1393042c-136424)}{3(70 c+29)}q^4
                      +O(q^5). \nn
\eea
Compared to (\ref{onepointfunctions}), there are corrections starting from the order $q^3$. As shown in Appendix~\ref{AppBasics}, on torus we have
\be
\lag W \rag_\mT=0.
\ee
The next primary operator is $\mE$ (\ref{E}) at level 6, and its leading contribution to the R\'enyi entropy is (\ref{e37}). This is beyond our consideration.
From (\ref{Snexpansion}), we get the R\'enyi entropy in expansion of $1/c$
\be
S_{n}^{(2,3)}=S_n^\cL + S_{n,(2)}^\NL+S_{n,(3)}^\NL+\cdots,
\ee
with the leading part $S_n^\cL$ equaling $S_n^\cl$ (\ref{Sncl}), the next-to-leading part $S_{n,(2)}^\NL$ equaling  $S_{n,(2)}^\oloop$ (\ref{Sn2oloop}), and the next-to-leading part $S_{n,(3)}^\NL$ equaling $S_{n,(3)}^\oloop$ (\ref{Sn3oloop}).

\section{$\mN\textbf{=(1,1)}$ SCFT}\label{SecSUSY}

In an $\mN=(1,1)$ SCFT, we have to consider the operators $G$ and $\bar G$, besides the stress tensor $T$, $\bar T$.
The $\mN=(1,1)$ SCFT is dual to the $\mN=1$ supergravity (SUGRA) in the AdS$_3$ spacetime.
The low temperature expansion of the additional R\'enyi entropy has been calculated on both SUGRA and SCFT sides \cite{Zhang:2015hoa}, and we further expand the result by the length of the interval
\bea \label{Sn32oloop}
&& S_{n,(3/2)}^\oloop= \Big( \frac{\pi^2(n+1)}{n}q^{3/2}
                            +\frac{5\pi^2(n+1)}{3n}q^{5/2}
                            +O(q^3) \Big)\ell^2
                      +\Big( -\frac{\pi^4(n+1) (13 n^2-17)}{60 n^3}q^{3/2}               \nn\\
&& \phantom{S_{n,(3/2)}^\oloop=}
                             -\frac{\pi^4(n+1) (47 n^2-51)}{36 n^3}q^{5/2}
                             +O(q^3)  \Big)\ell^4
                      +\Big( \frac{\pi^6(n+1) (205 n^4-614 n^2+457)}{7560 n^5}q^{3/2}    \\
&& \phantom{S_{n,(3/2)}^\oloop=}
                            +\frac{\pi^6(n+1) (4663 n^4-9610 n^2+5027)}{7560 n^5}q^{5/2}
                            +O(q^3) \Big)\ell^6 + O(\ell^8).                             \nn
\eea
On CFT side, we consider the torus partition function
\be
Z=q^{-\f{c}{24}}\bar q^{-\f{c}{24}}\prod_{k=0}^\inf\f{(1+q^{k+3/2})(1+\bar q^{k+3/2})}{(1-q^{k+2})(1-\bar q^{k+2})}.
\ee
From (\ref{T}) and (\ref{ABD}) we get the one-point functions
\bea
&& \lag T \rag_\mT = \frac{\pi^2 c}{6}-6 \pi^2 q^{3/2}-8 \pi^2 q^2-10 \pi^2 q^{5/2}+O(q^3) ,\nn\\
&& \lag \mA \rag_\mT = \frac{ \pi^4 c (5 c+22)}{180}
                      -2\pi^4(c-16) q^{3/2}
                      +\frac{8\pi^4(5 c+22)}{3} q^2
                      -\frac{2\pi^4(5 c-284)}{3} q^{5/2}
                      +O(q^3) ,\nn\\
&& \lag \mB \rag_\mT = -\frac{62\pi^6c}{525}
                      +\frac{72\pi^6}{25} q^{3/2}
                      +\frac{96\pi^6 (120 c+1)}{25} q^2
                      +696\pi^6q^{5/2}
                      +O(q^3) ,\\
&& \lag \mD \rag_\mT = \frac{\pi^6 c (2 c-1) (5 c+22) (7 c+68)}{216 (70 c+29)}
                      -\frac{\pi^6 (70 c^3-1903c^2+20054c+8104)}{2(70 c+29)}q^{3/2} \nn\\
&& \phantom{\lag \mD \rag_\mT =}
                      +\frac{22\pi^6 (2 c-1) (5 c+22) (7 c+68)}{3(70 c+29)}q^2
                      -\frac{\pi^6(350 c^3-38075 c^2+1065526 c+59624)}{6 (70 c+29)} q^{5/2}
                      +O(q^3).\nn
\eea
Since $G$ is fermionic, in the NS sector we have $\lag G \rag_\mT=0$. The next primary operator is $\mC$ (\ref{C}) at level 4, and its leading contribution to the R\'enyi entropy is (\ref{e39}). Similar to the case of CFT with $\mW(2,3)$ symmetry, this is beyond our consideration.
From (\ref{Snexpansion}), we get
\be
S_n^{(2,3/2)}=S_n^\cL + S_{n,(2)}^\NL+S_{n,(3/2)}^\NL+\cdots,
\ee
with the leading part $S_n^\cL$ equaling $S_n^\cl$ (\ref{Sncl}), the next-to-leading part $S_{n,(2)}^\NL$ equaling $S_{n,(2)}^\oloop$ (\ref{Sn2oloop}), and the next-to-leading part $S_{n,(3/2)}^\NL$ equaling $S_{n,(3/2)}^\oloop$ (\ref{Sn32oloop}).

\section{Conclusion}\label{SecConc}

In this paper we investigated the short interval expansion of the single interval R\'enyi entropy for a two-dimensional CFT on torus.
The length of the interval $\ell$ is small such that we could apply the OPE of the twist operators to reduce the computation to the one-point functions on the torus. Even though the prescription could be applied to any CFT, we focused on the large $c$ CFT dual to the AdS$_3$ gravity and we got the R\'enyi entropy to order $\ell^6$. We discussed the cases of vacuum conformal family, the CFT with $\mW(2,3)$ symmetry, and $\mN=(1,1)$ SCFT, and found consistent agreement with the existing results in all the cases. We discussed the case with chemical potential as well.

In the previous studies of the R\'enyi entropy in a large $c$ CFT on the torus \cite{Chen:2014unl,Chen:2015uia,Chen:2015kua}, there is no restriction on the length of the interval except that it should not be comparable with the size the torus , but one has to take the low or high temperature limit in order to expand the thermal density matrix appropriately. It is unwieldy to get the higher orders thermal corrections to the R\'enyi entropy. In the new prescription based on the short interval expansion, we have to take the small interval limit and only get the first few orders of interval length. However, the higher order thermal corrections, or finite size corrections, of the R\'enyi entropy can be got easily. In particular, we may read the $1/c$ correction quite easily, which is hard to find in the old treatment. Furthermore, we can study the case with nonvanishing chemical potential in the new prescription. The R\'enyi entropy in this case can be read easily. It would be nice to see if this can be reproduced on the gravity side.


\section*{Acknowledgments}

B.\,Chen would like to thank YITP for hospitality, where the project was finished.
B.\,Chen was in part supported by NSFC Grant No.\,11275010, No.\,11325522 and No.\,11335012.
J.-B.\,Wu and J.-j.\,Zhang were in part supported by NSFC Grant No.\,11222549 and No.\,11575202.
B.\,Chen, J.-B.\,Wu and J.-j.\,Zhang would also like to thank the participants of the advanced workshop ``Dark Energy and Fundamental Theory'' supported by the Special Fund for Theoretical Physics from NSFC with Grant No.\,11447613 for stimulating discussion.

\appendix

\section{Some basics of CFT}\label{AppBasics}

In this appendix we review some useful basics of the two-dimensional CFT. We mainly give some quasiprimary operators in the vacuum conformal family, and some primary operators in CFT $\mW(2,3)$ symmetry and $\mN=(1,1)$ SCFT.

The operators in the holomorphic sector of the vacuum conformal family can be written as quasiprimary operators and their derivatives. In level 2, one has the quasiprimary operator $T$, with the usual normalization $\a_T=\f{c}{2}$. In level 4, we have
\be \label{Adef}
\mA=(TT)-\f{3}{10}\p^2T, ~~ \a_{\mc A}=\f{c(5c+22)}{10}.
\ee
In level 6, we have
\bea \label{BDdef}
&& \mB=(\p T\p T)-\f{4}{5}(\p^2TT)-\f{1}{42}\p^4T, ~~
    \a_{\mc B}=\frac{36c (70 c+29)}{175},                                    \nn\\
&& \mD=(T(TT))-\f{9}{10}(\p^2TT)-\f{1}{28}\p^4 T +\f{93}{70c+29} \mc B,      \\
&& \a_{\mc D}=\frac{3 c (2 c-1) (5 c+22) (7 c+68)}{4 (70 c+29)}.             \nn
\eea
Under a general coordinate transformation $z\to f(z)$, we have
\bea \label{conftransf}
&& T(z)=f'^2 T(f)+\f{c}{12}s, ~~
   \mA(z)=f'^4\mA(f)+\f{5c+22}{30}s \Big( f'^2 T(f)+\f{c}{24}s \Big),                     \nn\\
&& \mB(z)= f'^6\mB(f)-\f{8}{5}f'^4s\mA(f)
          -\f{70c+29}{1050}f'^4s\p^2T(f)
          +\f{70c+29}{420}f'^2(f's'-2f''s)\p T(f)                                         \nn\\
&& \phantom{\mB(z)=}
          -\f{1}{1050}\lt( 28(5c+22)f'^2s^2+(70c+29)(f'^2s''-5f'f''s'+5f''^2s) \rt)T(f)   \\
&& \phantom{\mB(z)=}
          -\f{c}{50400}\lt( 744s^3+ (70c+29)(4ss''-5s'^2) \rt),                           \nn\\
&& \mD(z)=f'^6\mD(f)+\f{(2c-1)(7c+68)}{70c+29}s \Big( \f{5}{4} f'^4\mA(f)+\f{5c+22}{48}s \big( f'^2T(f)+\f{c}{36}s \big) \Big),   \nn
\eea
with the definition of Schwarzian derivative
\be
s(z)=\f{f'''(z)}{f'(z)}-\f32 \bigg( \f{f''(z)}{f'(z)} \bigg)^2.
\ee

For holomorphic sector of the CFT with $\mW(2,3)$ symmetry, we have the primary operator $W$ at level 3 with the usual normalization $\a_W=\f{c}{3}$.
All the operators constructed by $T$ and $W$ are of integer spins, and so for any such quasiprimary operator $\phi$ we have
\be
C_{\phi W \phi}=0.
\ee
So from (\ref{e34}), we conclude that on torus
\be
\lag W \rag_\mT=0.
\ee
At level 6 there is primary operator
\be \label{E}
\mE= (WW)-\f{1}{84}\p^4 T -\f{40}{9(5c+22)}\p^2\mA
    +\frac{215}{18(70 c+29)}\mathcal{B}
    -\frac{16 (191 c+22)}{3 (2 c-1) (5 c+22) (7 c+68)}\mathcal{D}.
\ee
There are the normalization and structure constant
\be
\a_\mE=C_{W\mE W}=\frac{4 c (c+2) (c+23) (5 c-4) (7 c+114)}{9 (2 c-1) (5 c+22) (7 c+68)},
\ee
from which we get in the large $c$ limit
\be
\f{(C_{W\mE W})^2}{\a_\mE\a_W^2} = 2 + O(1/c).
\ee
Using (\ref{AddTerm}), we have the term in R\'enyi entropy under the limit of small $\ell$, small $q$, and large c
\be \label{e37}
\ell^{12}b_{\mE\mE}\lag\mE\rag_\mT^2 \sim \ell^{12}q^6c^0.
\ee

For holomorphic sector of the $\mN=(1,1)$ SCFT, we have the primary operator $G$ at level 3/2 with the usual normalization $\a_G=\f{2c}{3}$.
Since $G$ is fermionic, in NS sector we have
\be
\lag G \rag_\mT=0.
\ee
The next primary operator is at level 4
\be \label{C}
\mC= (\p GG) - \f{3}{10}\p^2T - \f{17}{5c+22}\mA.
\ee
There are the normalization and structure constant
\be
\a_\mC=C_{G\mC G}=\frac{c (4 c+21) (10 c-7)}{6 (5 c+22)},
\ee
and in the large $c$ limit we get
\be
\f{(C_{G\mC G})^2}{\a_\mC\a_G^2} = 3 + O(1/c).
\ee
Under the limit of small $\ell$, small $q$, and large c, we have the term in R\'enyi entropy
\be \label{e39}
\ell^{8}b_{\mC\mC}\lag\mC\rag_\mT^2 \sim \ell^{8}q^3 c^0.
\ee

\section{Low temperature expansion of one-point functions}\label{AppLow}

In this appendix we reproduce the low temperature expansion of the one-point functions (\ref{onepointfunctions}) in a different method. We use the strategy that is introduced in the end of Section~\ref{SecRenyi}. In the calculations we need the normalization factors
\be
\a_T=\frac{c}{2}, ~~ \a_{\p T}=2c, ~~ \a_{\p^2 T}=20c, ~~ \a_\mA=\frac{c(5c+22)}{10},
\ee
and the structure constants
\bea
&& C_{TTT}=c, ~~
   C_{TT\mA}=\frac{c(5c+22)}{10}, ~~
   C_{TT\mB}=-\frac{2c(70 c+29)}{35},                           \nn\\
&& C_{TT\mD}=0, ~~
   C_{T\mA\mA}=\frac{2 c (5 c+22)}{5}, ~~
   C_{\mA\mA\mA}=\frac{c (5 c+22) (5 c+64)}{25},                \\
&& C_{\mA\mA\mB}=-\frac{4 c (5 c+22) (14 c+73)}{35}, ~~
   C_{\mA\mA\mD}=\frac{6 c (2 c-1) (5 c+22) (7 c+68)}{70 c+29}. \nn
\eea
We have the CFT on a cylinder with coordinate $z$, spatial period $L=1$, and the holomorphic thermal density matrix (\ref{rvac}). For $T$, we have the expectation value
\be
\lag T(z) \rag_\mT=\f{\tr \big[\r_\vac T(z)\big]}{\tr\r_\vac}=\f{ \lag0|T(z)|0\rag
                                                                 +\f{q^2\lag T|T(z)|T\rag}{\a_T}
                                                                 +\f{q^3\lag\p T|T(z)|\p T\rag}{\a_{\p T}}
                                                                 +\f{q^4\lag{\p^2T}|T(z)|{\p^2T}\rag}{\a_{\p^2T}}
                                                                 +\f{q^4\lag\mA|T(z)|\mA\rag}{\a_\mA}                                                                 +O(q^5)}{1+q^2+q^3+2q^4+O(q^5)}.
\ee
The three-point correlation functions on the right side can be calculated by mapping the cylinder to a complex plane $\mC$ with coordinate $f(z)=\ep^{2\pi\ii z}$. For example, we have
\be
\lag \p T |T(z)|\p T\rag=f'^2 \lag \p T(\inf)T(f)\p T(0)\rag_\mC+\f{c}{24}\lag \p T(\inf)\p T(0)\rag_\mC=\f{\pi^2c(c-72)}{3}.
\ee
Similarly, we can calculate other three-point correlation functions and then get the low temperature expansion of $\lag T \rag_\mT$. In the way method we can get $\lag \mA \rag_\mT$, $\lag \mB \rag_\mT$, and $\lag \mD \rag_\mT$ to the order $q^4$. The results are the same with (\ref{onepointfunctions}).

\providecommand{\href}[2]{#2}\begingroup\raggedright\endgroup


\end{document}